\documentclass[review]{elsarticle}
\usepackage[utf8]{inputenc}
\usepackage{xcolor}
\usepackage{amsmath}
\usepackage{bm}
\usepackage{hyperref}

\usepackage{subcaption}

\journal{Arxiv}

\begin{document}

\begin{frontmatter}

\title{The aerodynamic response of a pre-stressed elastically nonlinear aileron}

\author[1]{Giovanni Corsi\corref{cor1}}
\ead{giovanni.corsi@uniroma1.it}
\author[2]{Francesco Battista}
\ead{francesco.battista@uniroma1.it}
\author[2]{Paolo Gualtieri}
\ead{paolo.gualtieri@uniroma1.it}
\author[1]{Stefano Vidoli}
\ead{stefano.vidoli@uniroma1.it}
\cortext[cor1]{Corresponding author}
\address[1]{Dipartimento di Ingegneria Strutturale e Geotecnica,
Sapienza Università di Roma, Via Eudossiana, 18, 00184 Roma, Italy}

\address[2]{Dipartimento di Ingegneria Meccanica e Aerospaziale, 
Sapienza Università di Roma, Via Eudossiana, 18, 00184 Roma, Italy}

\begin{abstract}
    We study the aerodynamic response of a pre-stressed curved aileron. Whilst the fluid flow is standard (high Reynolds air flow undisturbed at infinity), the structure is designed to have a peculiar nonlinear behavior. Specifically, the aileron has only one stable equilibrium when the external forces are vanishing, but it is bistable when relevant aerodynamic loads are applied.
    Hence, for sufficiently high fluid velocities, another equilibrium branch is possible.  We test a prototype of such an aileron in a wind tunnel; a sudden change (snap) of the shell configuration is observed when the fluid velocity exceeds a critical threshold: the snapped configuration is characterized by sensibly lower drag. However, when the velocity is reduced to zero, the structure recovers its initial shape. A similar nonlinear behavior can have important applications for drag reduction strategies. 
\end{abstract}


\end{frontmatter}

\section{Introduction}
A promising research paradigm in the recent years has been the use of smart (or 
responsive) materials and structures, as enhancements of the functions of 
machines, or, in some cases, in order to replicate some of these functions 
altogether \cite{matmachJames2005}. Fields where this paradigm has found many 
applications have been, among others, robotics \cite{matAsMach2020} and 
aeronautics \cite{sun2016morphing,loewy1997recent, gomez2011morphing}. The 
functions that these materials allow can be divided in two broad categories 
\cite{hu2015buckling}: energy-related, concerning applications such as energy 
harvesting, or motion-related. One particular advantage is that these 
functions can usually be performed with reduced complexity of the final system, 
in that there's less need of actuating machinery, or complex control systems (if 
passive actuation is used). These applications can be obtained by taking 
advantage of the characteristics of the materials, such as memory shape alloys 
\cite{barbarino2014review}, hydrogels, elastomers, or polymers. An alternative 
is that of exploiting mechanical processes, such as occurrence of instabilities. 
A great example of this is buckling \cite{hu2015buckling}, an elastic 
instability that gives place to high deformations and sudden energy release, 
and can be exploited for both energy harvesting applications, motility, or 
design for actuators, among others.
In order to induce such mechanical processes, an interesting approach is that 
of taking advantage of both the characteristics of the 
material and of the geometry of the structure. For example, thin shells, after 
an incompatible elastic strain (as obtained, for example, by inelastic 
deformation during the production process), can result in spontaneous 
curvature and multi-stable behavior, which can be actuated in order to obtain 
complex motions or yet again for energy harvesting applications. 
In this work, we consider the case of a morphing shell with application in the 
context of aerodynamics. These shells are such that they have a pseudo-conical 
natural configuration, and, when clamped on one 
of the sides, tend to assume a curved shape, with possible bi-stable behavior 
when the initial curvature is chosen accordingly. The use of bistable structures 
has been considered already in aeronautics \cite{mattioni2006snap, MATTIONI2009151}, 
for example, for alleviation of excessive aerodynamic loads 
\cite{arrieta2014passive}. In this regard, we note some of the advantages of 
our approach, and of the particular features of the structure presented: 
the goal is to exploit the aerodynamic drag as actuation, inducing the bistable 
behavior after a critical loading threshold, obtaining sudden shape changes 
with consequent drag reduction. The
system can be tuned to any critical load, and to the desired multi-stable 
behavior by simply changing the starting geometry before clamping, and thus is 
very versatile compared to other usual applications, when the material and 
geometry both have to be fine tuned for each particular application. 
Furthermore, since actuation is due to the aerodynamic load alone, there's no 
need for complex actuations, such as heating, fluid supply, or chemical 
reactions. The dependence of the behavior on the load 
also means that the snap-through of the structure is trivially reversible: when 
the load decreases (that is, deceleration of the external flow) under a certain 
threshold, the structure snaps back to original configuration. This makes 
the shell an excellent example of smart structure in the context of 
aeroelasticity. It should be emphasized is that the structure is not bistable 
if a load is not applied: it will be shown that applying a load at the tip, 
after the shell is clamped at one of the sides, and increasing the force 
progressively, such a morphing shell can deform and snap-through, due to 
the emergence of a second stable configuration after a certain load threshold. 
This capability with a simple loading is easy to demonstrate. 
The purpose of this work is to demonstrate the feasibility of this approach in 
a more complex application, that is when the structure is under a realistic 
aerodynamic load. Such a load can be generated, for example, when the shell 
is mounted on a vehicle that moves at increasing speed: the pressure drag 
generated acts as an actuating force and could induce, in suitable conditions, 
snap-through. It remains to prove that with this complex and time dependent 
load, the desired behavior is still achieved. 
In order to do this, we study the problem both experimentally and numerically. 
The first part of the study is 
experimental: a prototype of this shell, fabricated with lamina 
of composite material, was studied in the wind tunnel, subject to increasing 
aerodynamic loads generated by an external flow. 
The numerical study is a simulation of the aeroelastic problem with a quasi-static 
approach: we simulate the drag due to an external flow, solving the set of 
$3D$ Navier-Stokes equations, accelerating from idle to a 
velocity that causes snap-through, and determine the response of the structure in 
incremental steps, as will be clarified below. In this way, 
we fully explore the application of a promising smart structure in the  
aeroelastic context. Such a behavior can have many interesting uses, such 
as the case, already mentioned, of load alleviation over a certain drag 
threshold, complex shape changes, as well as energy harvesting. Moreover, 
the results will confirm 
the feasibility of using reduced models for the description of the 
(non-linear) structural response to aerodynamic loads. In section 
\ref{sec:probstat}, the problem is stated in more detail. In 
section \ref{sec:experimental_evidence} we report the features of the 
structure, and the setup and results of the experimental study. In 
section \ref{sec:numerics} we describe the numerical study and we report 
numerical results, and their comparison with the experimental ones. We draw 
conclusions and perspectives on possible future work in section 
\ref{sec:conclusions}.

\section{Problem statement}
\label{sec:probstat}
\begin{figure}
    \centering
  \includegraphics[width=\linewidth]{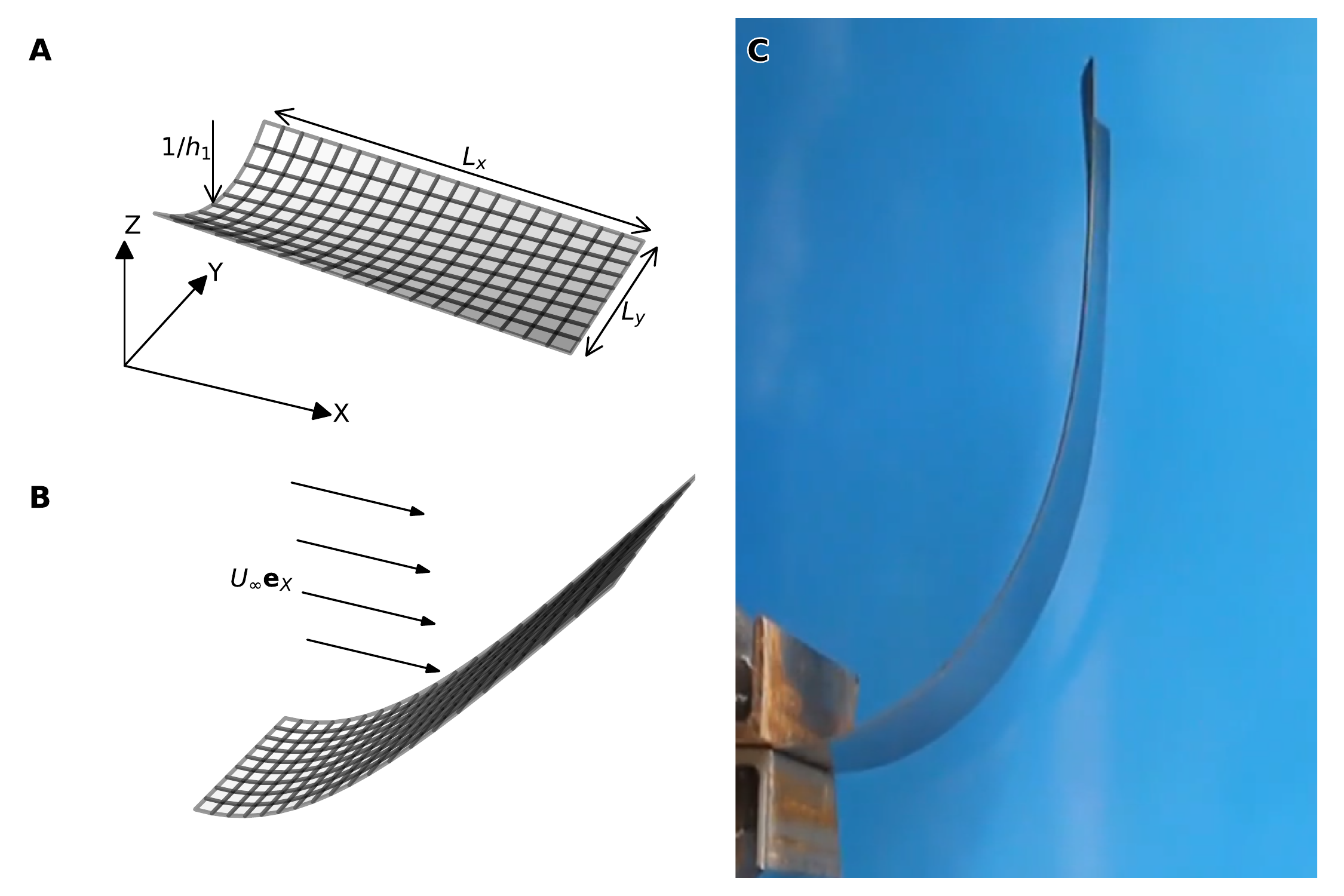}
    \caption{(A) natural configuration of the shell. (B) equilibrium shape 
        of the shell after clamping at the initially curved side. Incoming flow 
        velocity $\bm{u}_{\infty}$ aligned with the $X$ direction, is 
        represented. (C) snapshot of the prototype used for the experiments,
        attaining a curved shape after clamping.}
\label{fig:probstat}
\end{figure}

The structure studied is a pre-stressed shell made of composite laminate shown in Fig.~\ref{fig:probstat} (more details on the material are 
given in section \ref{subsec:mat_methods}).
In its natural stress-free configuration, the shell has rectangular planform with sides $L_x$ and $L_y$; one side of the shell has curvature 
$h_1$, while the other is flat. We refer the interested reader to \cite{BRUNETTI201647, brunetti2018bistability} where a similar class of 
shells subject to clamped boundary conditions was extensively studied.
When, the curved side is clamped, the shell equilibrium shape is 
changed into the curved profile shown in figure \ref{fig:probstat}B. This equilibrium shape is associated with a relevant stress field which is the principal responsible of the nonlinear elastic behavior of such a structure.

In the case studied, the natural configuration (namely $h_1$) is chosen such that the 
shell is monostable after clamping, but can become bistable after application 
of external loads. As anticipated, this is the novelty of our approach,
since bistability is not induced by means of complex phenomena or actuations, 
but is rather obtained, in a reversible manner, only as a result of external 
forces, in certain loading conditions.

\begin{figure}
    \begin{center}
  \includegraphics[width=\linewidth]{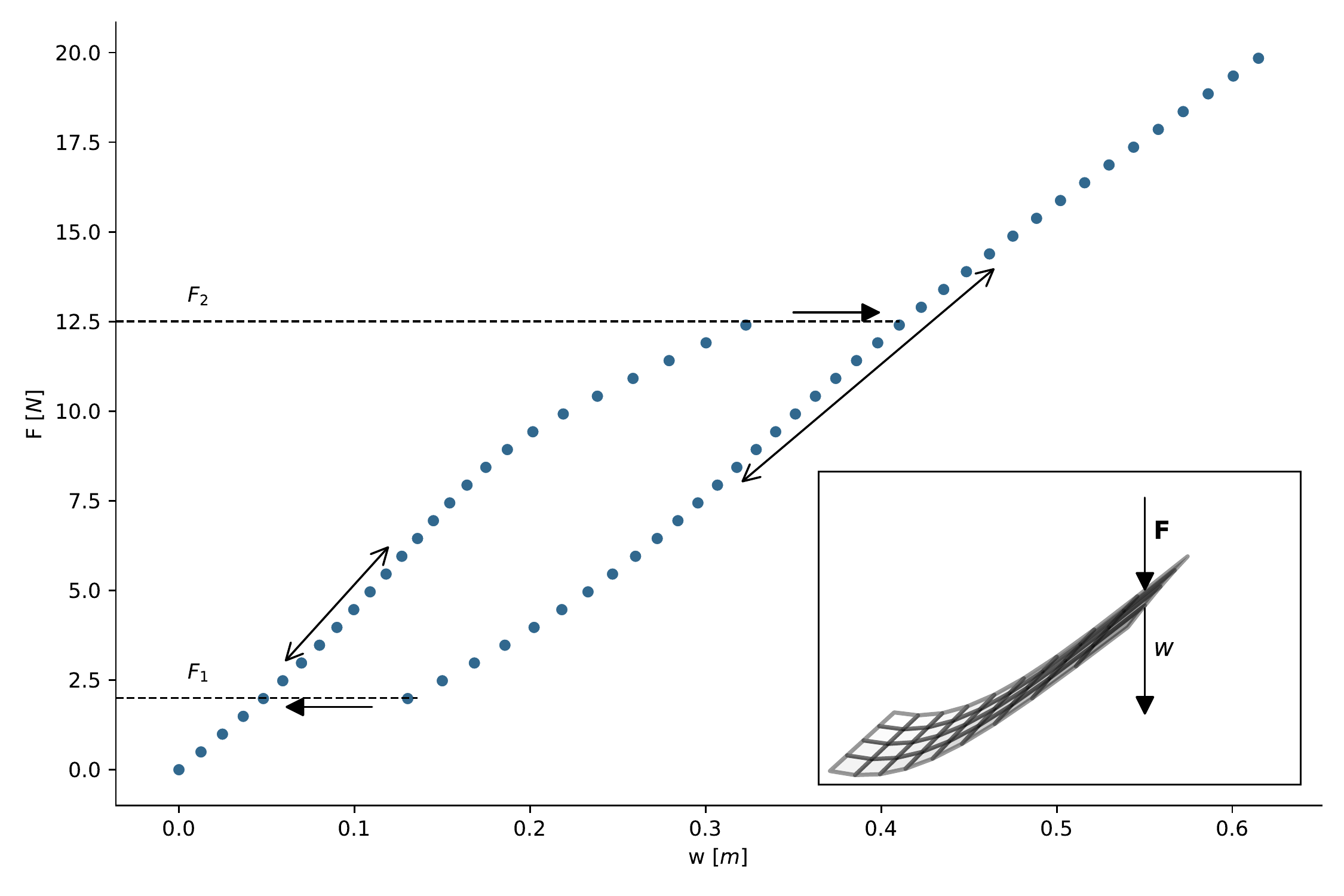}
    \caption{Displacement of the tip of the shell, in the clamped 
        configuration, as function of the value of a follower load, applied 
        at the tip as well. Possible movements along the solution branches, 
        and thresholds at which jumps between the two can be attained, are 
        depicted.}
\label{fig:follower_load}
    \end{center}
\end{figure}

Hence, when the applied external forces are vanishing, the curved equilibrium shape shown in Fig.~\ref{fig:probstat}B is the only stable equilibrium. However, the presence of forces could alter the shell stability behavior. For instance, we report in Fig.~\ref{fig:follower_load} the shell behavior under the application of a dead load $F$ at its tip. 

As such a force is progressively increased we compute all the possible stable equilibria of the shell and, in particular, their tip displacements shown in the horizontal axis of Fig.~\ref{fig:follower_load}. 
To this aim, we find all the minima of a reduced form of the elastic elastic energy of the shell as obtained in \cite{VIDOLI20131241}; deatils of this derivation are reported in \ref{sec:appendix}.
The energy has a polynomial expression and for a simple follower load, the 
system to be solved is simple enough that global minima can be found, with 
standard numerical tools, for each value of applied force. In figure 
\ref{fig:follower_load}, it can be seen that tho different solution branches 
are present: starting from the initial clamped configuration, for a force of 
value $F<F_2$, the behavior is hyperelastic. If the load is decreased, the 
solution would follow the same branch in the opposite direction. However, 
after a certain loading threshold $F_1$ the structure becomes bistable, and 
if the load is high enough, $F>F_2$, only the branch that emerges for higher 
force values remains. The structure will snap to the new branch, which is 
also stable and therefore can be followed by increasing or decreasing the 
load, again with an hyperelastic behavior. Only when the force decreases 
under the bistability threshold value, $F<F_1$, does the structure snap back 
to the first branch, which is the only one stable for low values of $F$. It 
can be readily noted that the two thresholds are different, $F_1 < F_2$. 
Therefore, the structure exhibits an hysteresis in the
loop obtained by loading and then releasing at the tip, if a certain threshold 
is exceeded. Again, it should be emphasized that this behavior was obtained 
through forcing alone, and no other means of actuation were necessary.

We investigate the behavior of such a structure when it is invested by a fluid flow in the axial direction with 
undisturbed velocity $\big(U_{\infty}$. 


 This 
experiment serves as a motivation for the rest of the work: the follower force 
at the tip is will be substituted with an aerodynamic load, as shown already 
in figure \ref{fig:probstat}. The action of the flow will be that 
of forcing the structure to open into a more streamlined shape. The flow 
velocity $U_{\infty}$ is supposed to increase in a quasi-static fashion, up to 
a threshold when multistable behavior occurs, and is subsequently decreased 
back to zero. This loop is the actuation that we want to investigate. The 
fluid of the external flow is supposed to be air, since we study the shell 
with the goal of showing its suitability for applications in the automotive 
or aerospace field, where the induced bistability could be exploited, e.g., 
for drag reduction.

\section{Experimental evidence}
\label{sec:experimental_evidence}
\begin{figure}
\centering
  \includegraphics[width=\linewidth]{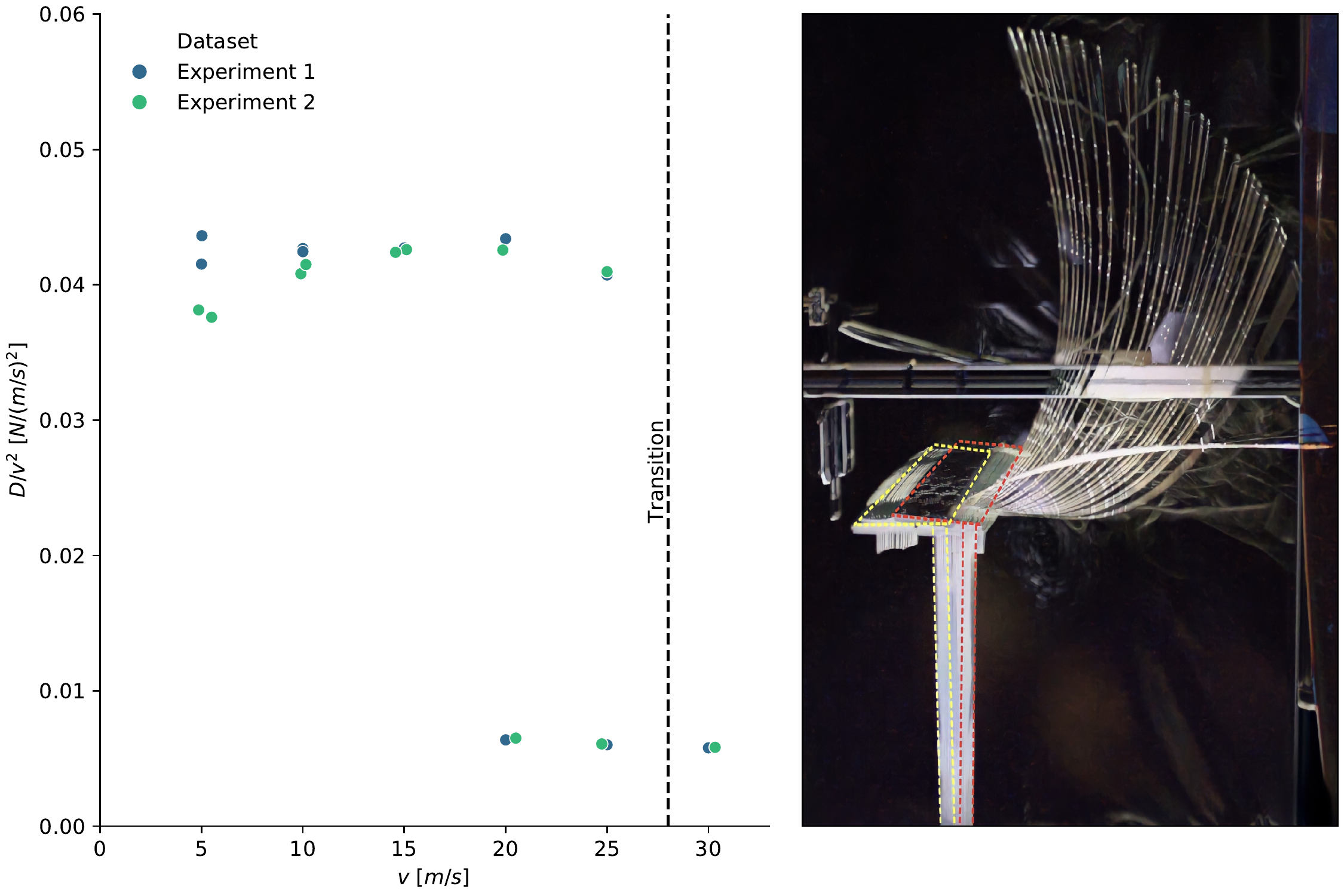}
    \caption{Left: experimental results in terms of measured drag forces, as 
        function of the external flow velocity. Right: superposition of 
        snapshots of an experimental run in wind tunnel, during an 
        acceleration of the external flow until the shell snaps to the other 
        stable configuration.}
\label{fig:exp}
\end{figure}

\subsection{Materials and methods}
\label{subsec:mat_methods}
The constitutive material of the shell studied in this work is a composite 
laminate, realized with the stacking of 8 layers, each an unidirectional  
carbon ply TenCate TC275-1 Epoxy Resin System, cured at $135 ^{\circ}C$ (see 
\cite{brunetti2018bistability} for a more detailed description of this 
composite shell). The stacking sequence of the layers, for a chosen 
$\alpha = 45^{\circ}$, is antisymmetric:

\begin{equation}
    [-\alpha / +\alpha / +\alpha / -\alpha / +\alpha / -\alpha / -\alpha / +\alpha],
\end{equation}
The stiffness matrices of the composite laminate, needed for the derivation 
of the elastic model (see equation (\ref{eq:constitutive_equations})) are: 

\begin{equation}
    \frac{\bm{A}}{A_{11}} = \frac{\bm{D}}{D_{11}} = \begin{pmatrix}
        1 & \nu & 0\\
        \nu & \beta & 0\\
        0 & 0 & \gamma
    \end{pmatrix}, \qquad \bm{B} = \bm{0}.
    \label{eq:stiffness_matrices}
\end{equation}
For the material use it holds $\beta = 1.0$, $\nu = 0.851$, $\gamma = 0.858$, 
$A_{11} = 4.59 \cdot 10^7 \frac{N}{m}$ and $D_{11} = 3.82 Nm$. The starting 
curvatures are $h_1 = \frac{1}{r_1}$ and $h_2 = \frac{1}{r_2}$, with 
$r_1 = 0.064 m$ and $r_2 = \infty$, meaning that only the side to be clamped 
is curved, while the other is flat. The final thickness of the laminate is 
$1mm$ ($0.125mm$ for each layer), while the dimensions of the natural 
configuration are $L_x = 0.33m$ and $L_y = 0.15m$ (see figure 
\ref{fig:probstat}).

\subsubsection{Experimental setup}
An experimental campaign was run in the wind tunnel facility at the Fluid 
Mechanics Lab at Sapienza University of Rome. The wing was supported in such 
a way that the structure is the only obstacle in an otherwise undisturbed 
flow. A view of the support of the shell is given in figure \ref{fig:exp} 
right: the side to be clamped is closed between two plates, and four screws 
are inserted and tightened to pull the plates together. The screws are 
positioned laterally with respect to the shell, that is, no holes are made on 
the structure. Two small layers of Styrofoam protect the shell avoiding direct 
contact with the plates. This clamping therefore is not perfect, in that the 
shell has a small, non-zero curvature within the two plates. This can be 
observed also from the fact that the two clamping plates are slightly curved: 
in correspondence of the screws, they are closed together, while some distance 
arises in-between due to the presence of the plate. Between each experimental 
run, this distance is measured in order to assure at least that it is always 
the same value. A perfect clamping would cause the risk of permanently deform 
the shell, preventing he reproduction of the experiment. The support of the 
wing was connected to a load cell 
in order to measure the drag force. During each experimental run, the velocity 
of the flow was varied in a controlled 
manner in small increments, from $0$ up to around $30 m/s$. The increase was 
stopped after snap-through was observed, and was slow enough in order to have 
a loading as close as possible to quasi-static. Once the maximum velocity was 
reached, a slow deceleration was imposed to get back to the starting velocity 
value. Several measurements of this acceleration-deceleration loop have been 
performed, paying particular attention to the threshold velocity values and 
to the reproducibility of the acceleration-deceleration loop response of the 
structure. At high velocities, close to the snap-through threshold, the 
support showed some degree of deformation. This can be observed in figure 
\ref{fig:exp} right: before the snap transition, the support has rotated 
slightly in direction of the flow. Once the shell snaps to the other stable 
configuration, the rotation vanishes. From the series of snapshots of the 
experiments, the maximum rotation angle is estimated to be around $3$ degrees. 

\subsection{Results}
\label{subsec:results}
In the experiments, velocity was gradually increased from rest until a 
snap-through was observed, then slowly decreased back to the starting value. 
In figure \ref{fig:exp}, we report the experimental measurements of drag on 
the shell during this loop. Several measurements were performed in 
order to highlight the reproducibility of the phenomena.
From the experimental drag results, it is clear that a snap-through occurs for 
a value of incoming flow velocity of about $29 m/s$. When this happens, the 
drag value measured on the structure falls, since the new 
stable equilibria is an almost flat configuration. This new configuration is 
maintained even when velocity is then decreased, gradually stopping the flow 
of the wind tunnel. This implies that the equilibrium state is moving along a 
stable branch different from the initial one. Only when velocity is 
considerably lower, around $15-17 m/s$, does the structure snap-back to the 
original stable branch. This transition is spontaneous and requires no external 
intervention. This proves that the structure bi-stability can be 
triggered by a force of the kind generated by aerodynamic drag. At both 
transitions between stable branches, significant oscillations are observed: 
this is indicative that the stiffness of the structure is decreasing to very 
low values, and is typical of a snap-through 
scenario; when the structure snaps from one stable configuration to the other, 
it can be thought to have zero-stiffness. The oscillations are not shown in 
figure \ref{fig:exp} because measurements during this phase are difficult to 
perform accurately. From figure \ref{fig:exp} it can also be noted that the 
transition loop can be reproduced very accurately in different measurements.

\section{Numerical evidence}
\label{sec:numerics}

With the goal of developing a numerical model capable of reproducing the 
experimental behavior, two different systems will be coupled: a reduced 
nonlinear elastic model for the shell, and a set of $3D$ Navier-Stokes 
equations. The fluid-structure coupling is approximated as quasi-static: 
indeed in the experiment the flow 
velocity varies slowly, and so does the structural deformation with the 
exception of the snap-through. Therefore, we don't solve for the dynamic 
strong coupling that would result from the full the aeroelastic problem, as 
detailed below. The reduced shell model, derived in \ref{sec:appendix}, yields 
an elastic energy functional that is polynomial in three lagrangian 
coordinates 

\begin{equation}
    \mathcal{E}_{el} = \hat{\epsilon}\left( \bm{d}(q_1, q_3, q_4) \right),
    \label{eq:elastic_energy}
\end{equation}
where $\bm{d}$ is the displacement field with respect to the initial (after 
clamping) configuration. The three coordinates (see \ref{sec:appendix} for 
their definition) have an immediate 
interpretation: $q_1$ is the curvature of the shell centerline at the point of 
clamping $x=0$, $q_3$ is the variation of curvature along the centerline, and 
$q_4$ is the curvature in the direction perpendicular to the centerline. To 
obtain the total energy, one has to subtract the work of the external forces, 
defined as:

\begin{equation}
    F_j = \int_{S_0}\bm{F}_v \cdot \frac{\partial \bm{d}}{\partial q_j} \text{d}S_0,
    \qquad j = 1,3,4
    \label{eq:gen_forces}
\end{equation}
where $\bm{F}_v$ is the vector field of the external forces acting on the 
surface of the shell.  
When solving the problem for a simple (or absent) forcing, such as the example 
in section \ref{sec:probstat} (figure \ref{fig:follower_load} in particular), 
the system of equations is simple enough that one can find all minima by 
using standard routines for root finding of polynomial systems. In particular, 
we use \textit{Mathematica} \cite{mathematica} and its \textit{NSolve} routine. 
The gradient of the total energy will indeed be polynomial, and the 
problem reduces to finding its roots. From these solutions then the minima of 
the potential energy are identified, since the hessian can be calculated 
directly as well. This procedure yields the global stability landscape, and 
allows to solve for all solution branches at the same time. On the other hand, 
when the forcing is given by the interaction with a fluid, the force field is 
more complex, since now $\bm{F}_v=-p\bm{n}$, with $p$ pressure of the fluid, 
given by the solution of the external flow equations, and $\bm{n}$ the normal 
to the shell surface. In this case one cannot hope to solve the problem with 
the direct method above. We resort to finding local minima iteratively. 
Starting from a known solution value $q_j^n$, $j = 1,3,4$, and knowing the 
fluid force $\bm{F}_v = -p^{n+1}\bm{n}$, the problem to solve is:

\begin{equation}
    \begin{aligned}
        q_j^{n+1} &= \min_{q_j} \mathcal{E}(q_j, q_{j}^n), \qquad j=1,3,4 \\
        \mathcal{E}(q_j, q_{j}^n) &= \hat{\epsilon}\left( \bm{d}(q_1, q_3, q_4) \right) - 
        \Sigma_j F_j(q_{j}^n)\cdot(q_j-q_{j}^n).
    \end{aligned}
    \label{eq:fluid_prob_pot}
\end{equation}
Several approximations have been done: in order to obtain an expression of the 
work in closed form, the force is supposed to be constant over the step, and 
is calculated for the shell configuration (on which the lagrangian forces 
depend implicitly) at the known initial solution $q_{j}^n$. This is the 
simplest approximation and is expected to work if the steps are small enough.
Our quasi-static solution cycle is thus composed by the following sub-steps:

\begin{itemize}
    \item start from the previous known solution, for a given state 
        $U_{\infty}^n$, $p^n$ of the fluid and $q_{j}^n$ for the solid. If at 
        beginning of simulation, $U_{\infty}^0 = 0 = p^0$, find minimum for zero 
        forcing to obtain starting $q_{j}^0$
    \item generate geometry of the shell given lagrangian coordinates $q_{j}^n$ 
        (see \ref{sec:appendix}), or the starting ones $q_j^0$ 
    \item given value of external velocity $U_{\infty}^{n+1}$ at the new step, 
        solve the external flow problem (detailed below) and calculate the new 
        pressure field $p^{n+1}$
    \item calculate lagrangian forces from equation (\ref{eq:gen_forces}) and 
        solve problem (\ref{eq:fluid_prob_pot}) to calculate the new values
        $q_j^{n+1}$
    \item find new local solution of the system (\ref{eq:etot_der_0}) in 
        terms of the values of $q_1$, $q_3$, $q_4$, using as starting point 
        for root search the previous known solution
    \item move to next step
\end{itemize}
The value of external flow velocity $U_{\infty}$ ranges in a loop, in small 
increments, from a zero starting value to a maximum value, taken from the 
experiments, and then back to the starting one. This allows us to simulate a 
loop of acceleration and subsequent deceleration of the flow, which is the 
working condition in which the structure is to be studied. To perform the 
local minimization step, we use the \textit{scipy} library.

\subsection{Navier-Stokes flow simulations}
\label{subsec:openfoam}
The external flow problem is modelled by the steady Navier-Stokes 
equations. To solve these at each step, we use the \textit{OpenFOAM} library. 
The mesh of the fluid domain is generated with the utility $snappyHexMesh$. 
The domain consists of a rectangular space both in the $xy$ and $xz$ planes, 
with a hole at the center, which coincides with the boundary of the structure 
in the configuration being currently studied. The center of the clamped side 
of the shell is positioned at the origin. The positioning of the solid surface, 
far from any other boundary, is consistent with the experimental set-up. The 
domain is large enough that the condition of undisturbed flow is attained far 
from the boundary of the shell. Keeping consistent with the coordinates in 
section \ref{sec:probstat}, the surface centerline develops in the plane $xz$, 
and has a weak curvature in the $xy$ plane. The incoming flow is oriented in 
the $x$ direction, therefore hitting the surface as a bluff body when this is 
curled after clamping, and would be tangent to it if the shell were completely 
opened into a flat shape, similar to what is observed in the experiments after 
the structure snaps to another stable solution branch. This geometry 
is defined by means of an STL surface, which is generated with a 
\textit{python} utility starting from a point sampling of the mid-surface, 
given from eq. (\ref{eq:mid_surface}). The solver utilized is 
\textit{simpleFoam}, suitable for computation of a stationary solution of the 
Navier-Stokes equations. In figure \ref{fig:follower_load}, the structure 
has a bistable behavior when applying a force at the tip, of magnitude of the 
order $10 \div 20 N$. To estimate roughly the magnitude of the incoming flow 
velocity, one can calculate which velocity would produce a comparable drag 
force over a flat plate with the same dimensions of the structure. This 
results in an estimate of Reynolds number of the order of $10^5 \div 10^6$, 
meaning that the flow is expected to be fully turbulent. At Reynolds numbers 
this high, we can assume that friction forces be negligible with respect to 
the pressure contribution. This justifies defining the fluid force field as 
$\bm{F}_v = -p\bm{n}$ in equation \ref{eq:gen_forces}. The $k-\varepsilon$ 
\textit{RANS} turbulence model is used for modeling of turbulence. The boundary 
conditions are of imposed velocity at the inlet and on the solid surface 
(\textit{no-slip} in the latter case), while the \textit{zeroGradient}, that 
is, Neumann boundary condition is assigned at the outlet. At the top and bottom 
surfaces (w.r.t. the $z$ direction) a \textit{slip} condition is assigned. The 
two remaining surfaces are set as symmetry planes. As to the pressure, a fixed 
value of $0$ is set at the outlet. Turbulent wall functions corresponding to 
the turbulence model in use are set at the solid boundary, in order to model 
the boundary layer. The turbulence variables $k$ and $\varepsilon$ to be 
assigned at inlet are estimated with the following formula: 

\begin{equation}
    \begin{aligned}
        k_{\infty} &= \frac{3}{2}(I U_{\infty})^2,\\
        \varepsilon_{\infty} &= \frac{1}{L_y}(\frac{2}{3} k_{\infty})^{\frac{3}{2}},
    \end{aligned}
\end{equation}
where $U_{\infty}$ is the velocity far from the structure, assigned at the 
inlet, and in our case $I = 0.1$ and $L_y$, introduced in section 
\ref{sec:probstat}, is taken as reference length. Discretization of the 
Navier-Stokes equations is done with finite volumes, with 
second order accuracy for regular meshes. The discretization of the fluxes of 
all variables is done with an interpolation scheme that ensures stability 
(\textit{TVD}), and we employ schemes in their \textit{bounded} variant, that 
is, schemes with an additional term that penalizes the continuity 
(incompressibility) errors. This term, added to the discretized equations, is 
consistent in that it vanishes when convergence to a solution is reached. 

\subsection{Numerical results}

\begin{figure}
\centering
    \includegraphics[width=\linewidth]{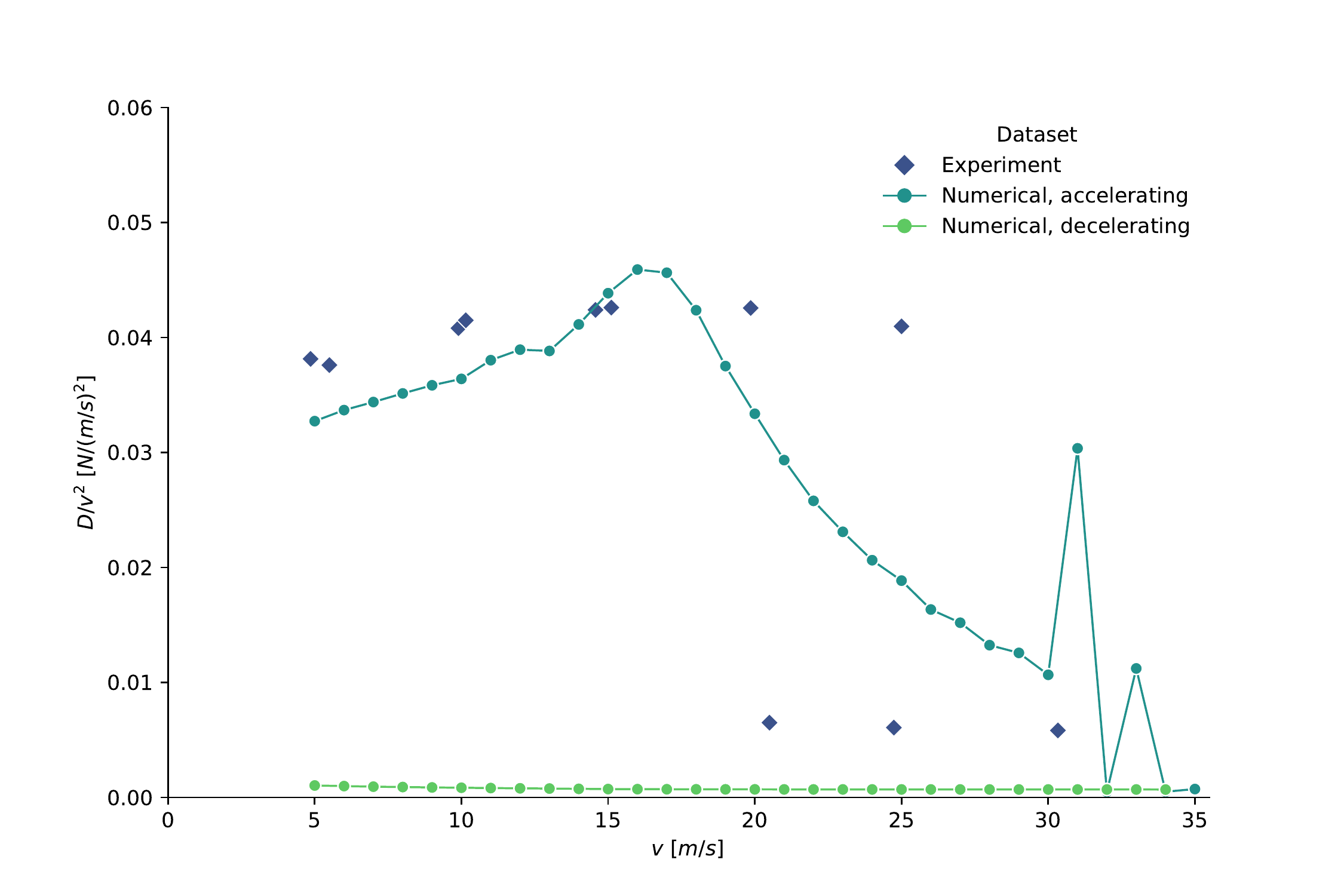}
    \caption{Comparison between numerical and experimental results in terms of 
        measured drag forces, as function of the external flow velocity.}
\label{fig:exp_comp}
\end{figure}
\begin{figure}
\centering
  \includegraphics[width=\linewidth]{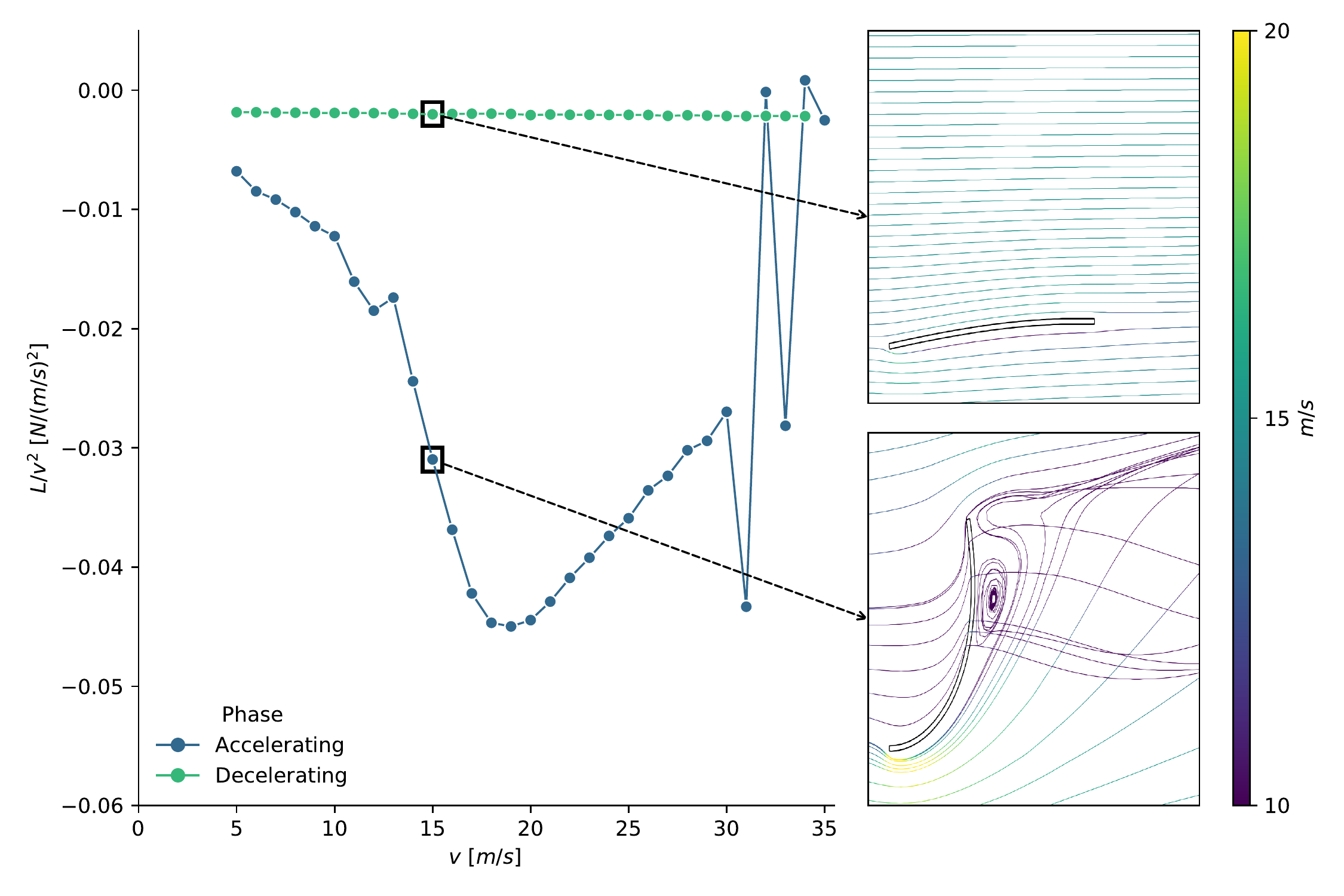}
    \caption{Numerical results, given as function of outer flow velocity. 
        Left: lift force (force in $z$ direction). Right: snapshots of 
        streamlines, from numerical simulations.}
    \label{fig:numlift}
\end{figure}

The results of the numerical model introduced in this section are compared 
with the experimental benchmark. In figure \ref{fig:exp_comp}, the drag force 
values resulting from an acceleration-deceleration loop of are reported, 
together with the experimental results. In figure \ref{fig:numlift} instead, 
the numerical results for the lift force are given, for which there is no 
experimental data (for our choice of coordinates, as in 
figure \ref{fig:probstat}, lift and drag are the force components $z$ and 
$x$, respectively). The solution procedure allows us to identify two branches 
of solutions. The first one originates from the equilibrium solution for zero 
external forcing. When velocity is increased past a certain threshold, which 
can be individuated around $29-30 m/s$, the system shows strong oscillations 
and stabilizes on a second branch, which is stable and corresponds to much 
lower values of drag. This second branch can then be followed increasing or 
decreasing the flow velocity, meaning that after the jump, the shell has an 
hyperelastic behavior on a different stable branch with respect to the 
starting one, as was seen in the example of figure \ref{fig:follower_load}. 
The branch remains stable even if velocity is decreased, gradually, back to 
the starting value. The second transition, from the second branch back to the 
starting one, is not observed numerically.
The initial numerical branch, that is, the solution branch obtained following 
starting from the equilibrium state for zero forcing, agrees qualitatively with 
the experimental results of the acceleration phase. Differences in the 
observed values of the drag, and in the reproduced behavior can be attributed 
to several factors: the transient contributions are neglected in the numerics, 
and the acceleration/deceleration of the flow in the experiments, even though 
the velocity variations were gradual, could be a factor in correspondence of 
the snap transitions between stable branches. Another factor, as reported in 
section \ref{sec:experimental_evidence}, is that in the experiments, the 
support over which the shell was fixed showed some degree of rotation at the 
higher velocities, as is highlighted in figure \ref{fig:exp} right. Furthermore, 
it should be noted that inaccuracies can also be attributed to the uncertainty 
in the measurement of the thickness of the shell: the bending stiffness 
scales with $t^3$, that is, with the third power of the thickness, thus small 
variations can result in non negligible stiffness changes.

\section{Conclusions}
\label{sec:conclusions}

Our contribution is two-fold: we presented an application of smart structure, 
where multi-stability is harnessed to induce interesting behavior. The shell 
considered here can be bistable when under external aerodynamic loading, 
proving valuable for drag reduction strategies. In particular, this 
bistable behavior, which is triggered by forces, is reversible and therefore no 
actuation is necessary. This behavior was first observed with a simple 
experiment applying a point follower load. With an experimental campaign in a 
wind tunnel, we then confirmed that an aerodynamic force can trigger the same 
bistable, and reversible, behavior. Moreover, we presented a numerical study 
of the aeroelastic problem, and proposed a strategy based on a reduced 
nonlinear elastic model and $3D$ steady Navier-Stokes fluid model. This 
approach has the advantage of being computationally inexpensive with respect 
to a fully coupled aeroelastic FSI simulation. The model is capable of predicting 
the existence of different solution branches, and reproduces the experimental 
data, for the accelerating flow phase, with an acceptable accuracy. The snap 
transition between stable branches at high flow velocities is predicted by the 
numerical model, while the snap back is not reproduced. We discussed the 
factors that have to be improved to improve the behavior reproduction of the 
numerical model. In particular, the effective clamping conditions in the 
experiments need to be assessed with care, as well as the possible 
deformations of the support for high aerodynamic loads. A further development 
could be assessing the performance of a more complicated solid model: the 
shallow shell model works well when the curvatures are low, which was not a 
limitation in our case since, due to our setup, the aeroelastic load forces 
the shell to assume a progressively more streamlined (and flat) shape. The 
solid model depends, through the ansatz of the displacement field 
(\ref{eq:ansatz_displacement}), on five lagrangian coordinates. The model 
could be enriched, adding polynomial terms to the ansatz, which could result 
in more accurate predictions.

\appendix

\section{Shell model}
\label{sec:appendix}

\subsection{Derivation of the nonlinear reduced model}
\label{subsec:solid_structure}
We refer to the design of multi-stable shells in \cite{BRUNETTI201647}, for 
which a model is given within the assumptions of the shallow shell 
approximation, starting from the Foppl von-K\'{a}rm\'{a}n theory. In this 
approximation, the following constitutive relations hold 
\begin{equation}
    \begin{aligned}
        \bm{t} &= \bm{A}(\bm{e} - \bm{f}) + \bm{B}(\bm{k}-\bm{h}),\\
        \bm{m} &= \bm{D}(\bm{k} - \bm{h}) + \bm{B}^T(\bm{e}-\bm{f}),
    \end{aligned}
    \label{eq:constitutive_equations}
\end{equation}
where, in vector notation, the membrane stresses $\bm{t}$, the bending moments 
$\bm{m}$, the membrane strains $\bm{e}$ and the curvatures $\bm{k}$ appear, 
with $\bm{f}$ and $\bm{h}$ stresses and curvatures of the natural configuration. 
$\bm{A}$ and $\bm{D}$ are the stiffness matrices, given in section 
\ref{subsec:mat_methods}.  The particular choice of starting natural configuration 
is what will induce the multi-stable behavior; no inelastic pre-stresses are 
applied, and therefore $\bm{f} = \bm{0}$, $\bm{h} \neq \bm{0}$.
The shell has a natural configuration which is 
pseudo-conic. Thus, the mathematical expression of its mid-surface will be of 
the form:
\begin{equation}
    \begin{aligned}
        S_0 &= \{(x,y,w_0(x,y)), 0\leq x\leq L_x, -L_y/2 \leq y \leq L_y/2\},\\
        w_0(x,y) &= \frac{y^2}{2}\left(h_1 +(h_2-h_1)\frac{x}{L_x}\right),
    \end{aligned}
    \label{eq:natconf}
\end{equation}
where $w$ is the displacement in the $z$ direction, and the shell in 
its flat configuration lies in the $xy$ plane. $h_1$ and $h_2$ determine the 
curvatures of the natural configuration. In the working configuration, the 
shell will be clamped at the side $x=0$, which is the side of curvature 
$h_1$.

\subsubsection{Reduced nonlinear model}
A model for the response of the shell under aerodynamic loads is needed: a 
low-dimensional model \cite{VIDOLI20131241} was proven to be effective in 
describing the the 
deformation of the shell under loading, as well as predicting eventual 
bifurcations and the post-critical behavior. More in detail, an ansatz is 
given for the expression of the transverse displacement field:
\begin{equation}
    w(x,y) = q_1 \frac{x^2}{2} + q_2 \frac{y^2}{2} + q_3\frac{x^3}{6} +
        q_4\frac{x^2y^2}{2} + q_5\frac{xy^2}{2},
    \label{eq:ansatz_displacement}
\end{equation}
from which we have the description of the natural configuration 
(\ref{eq:natconf}):

\begin{equation}
    q_1 = q_3 = q_4 = 0, \quad q_2 = h_1, \quad q_5 = \frac{h_2-h_1}{L_x}.
\end{equation}
The clamping boundary condition at the boundary $x=0$ results in the 
simplification $q_2 = q_5 = 0$, reducing the problem to the determination of 
just the three coefficients $q_1$, $q_3$, $q_4$. We now briefly recall the 
solution process, explained in detail in \cite{BRUNETTI201647, 
VIDOLI20131241}, of the solid problem: with the simplifications introduced, 
it is possible to solve the elastic problem for the equilibrium of the 
structure in closed form. The membrane strains indeed must satisfy the 
compatibility relation

\begin{equation}
    \varepsilon_{x,yy} + \varepsilon_{y,xx} -2 \varepsilon_{xy,xy} = 
        k_x k_y - k_{xy}^2,
\end{equation}
where in the subscript we indicate components followed, after a comma, by the 
eventual partial derivatives. The curvatures $k_x$, $k_y$ and $k_{xy}$, due 
to our model assumptions, depend only on the transverse displacement

\begin{equation}
    k_x = w_{,xx}, \quad k_y = w_{,yy}, \quad k_{xy} = w_{,xy},
\end{equation}
and will have a simple polynomial form, due to the ansatz 
(\ref{eq:ansatz_displacement}). Thus, one can rewrite the problem for the 
membrane stresses and strains in terms of the transverse displacement $w$. 
This problem will be linear in the data \cite{VIDOLI20131241}, and therefore 
it is possible to derive an expression of 
the membrane stress and strain fields which is a linear polynomial in the 
lagrangian unknowns $q_j$. Adding the bending contributions as well, the 
resulting elastic energy of the shell will have a polynomial expression 

\begin{equation}
    \mathcal{E}_{el} \simeq \hat{\epsilon}(q_1, q_3, q_4),
\end{equation}
The forcing term will be calculated from the fluid force field 
$\bm{F}_v$ due to the flow investing the structure

\begin{equation}
    F_j = \int_{S_0}\bm{F}_v \cdot \frac{\partial \bm{d}}{\partial q_j} \text{d}S_0,
    \label{eq:gen_forces_app}
\end{equation}
where the lagrangian vector field $\bm{d}$ represents the displacement of a 
point in $S_0$, and it is implicit that the force vector $\bm{F}_v$ has been 
pulled back to the reference lagrangian configuration. The new configuration 
given the forcing will be found among the solutions of 

\begin{equation}
    \frac{\partial \hat{\epsilon}}{\partial q_j} - F_j = 0, \qquad j = 1, 3, 4.
    \label{eq:etot_der_0}
\end{equation}
One of the advantages of the reduced model is that if no load is applied, or 
if the loading has a simple polynomial expression, solving 
(\ref{eq:etot_der_0}) becomes a matter of finding roots of a polynomial. 
Selecting the (real) solutions that represent a minimum of the energy, 
one can find all branches of equilibrium. If the problem is more complex due to 
non-trivial forcing, one has to resort to finding solutions of 
(\ref{eq:etot_der_0}) with local root finding routines, or, as in our case, to 
performing local minimization of an approximate potential energy.

\subsection{Reconstruction of the surface}
The mid-surface of the structure can be derived from the displacement field 
given by (\ref{eq:ansatz_displacement}), which yields a good approximation
in the shallow-shell regime. It is a sufficiently accurate approximation when 
the shell is close to it natural configuration. The shallow approximation can 
be used directly for both the curvatures and the displacements: the surface vector 
field will be 

\begin{equation}
    \bm{S}(x,y) = \big(\big. x, y, w(x,y)\big.\big),
    \label{eq:mid_surface}
\end{equation}
with $w(x,y)$ from (\ref{eq:ansatz_displacement}).

To obtain a surface shape that is valid for larger displacements, further 
away from the natural configuration, we cannot simply use eq. 
(\ref{eq:ansatz_displacement}), but need to use a shallow-shell 
approximation only in the curvatures; for example, the mid-axis of the shell, 
intended as the curve $w(x, 0)$, might be a circular arc. In this case, once 
the solution coordinates $q1$, $q3$, $q4$ are known, we reconstruct the surface 
as follows: from the expression of the centerline $w(x, 0)$, that of the angle of 
its tangent vector readily follows, integrating the curvatures: 

\begin{equation}
    \theta(s) = \int_0^s \frac{\partial^2 w(x,0)}{\partial x^2} \text{d}x,
\end{equation}
from which the reconstructed centerline is 
\begin{equation}
   \bm{p}(x) = \int_0^s \begin{pmatrix}
       \cos\theta(s)\\
       0\\
       \sin\theta(s)
   \end{pmatrix}\text{d}s.
   \label{eq:centerline}
\end{equation}

In our case, we assume that stretching of the centerline be negligible, and 
therefore $x$ and $s$ coincide. From the centerline, the rest of the (mid-) 
surface will be given by adding a parabolic term in $y$:

\begin{equation}
   \bm{S}(x, y) = \bm{p}(x) + \frac{\partial^2 w}{\partial y^2} \frac{y^2}{2}
   \begin{pmatrix}
       \cos \left( \theta(x)+\frac{\pi}{2} \right)\\
       0\\
       \sin \left( \theta(x)+\frac{\pi}{2} \right) 
   \end{pmatrix}.
   \label{eq:mid_surface}
\end{equation}
Given the reconstruction of the vector field $\bm{S}(x,y)$, we calculate the 
tangent and normal vectors as 

\begin{equation}
    \begin{aligned}
        \bm{g}_x &= \frac{\partial \bm S(x,y)}{\partial x},\\
        \bm{g}_y &= \frac{\partial \bm S(x,y)}{\partial y},\\
        \bm{n}(x,y) &= \frac{\bm{g}_x \times \bm{g}_y}{
            \lVert \bm{g}_x \times \bm{g}_y \rVert  }.
    \end{aligned}
    \label{eq:tangent_and_normal}
\end{equation}
All of the fields calculated so far 
implicitly depend on the three lagrangian coefficients $q_1$, $q_3$, $q_4$. The 
derivatives of the displacement $\frac{\partial \bm{d}(x,y)}{\partial q_j}$, 
needed in equation (\ref{eq:gen_forces_app}), will be given by differentiating 
the field $\bm{S}(x,y)$.

\bibliography{bibliography}

\end{document}